\documentclass[10pt, conference, a4paper]{IEEEtran}

\usepackage{amsmath}
\usepackage{booktabs}
\usepackage{verbatim}
\usepackage{graphicx}
\usepackage{balance} % Added to balance final page columns

\usepackage{tikz}
\usetikzlibrary{shapes.geometric, arrows.meta, positioning}
\usepackage{pgfplots}
\usepgfplotslibrary{groupplots}
\pgfplotsset{compat=1.18}

\begin{document}

\title{The Cognitive Circuit Breaker: A Systems Engineering Framework for Intrinsic AI Reliability}

\author{
    \IEEEauthorblockN{Jonathan Pan}
    \IEEEauthorblockA{
        Home Team Science and Technology Agency\\
        Singapore\\
        Jonathan\_Pan@htx.gov.sg
    }
}

\maketitle

\begin{abstract}
As Large Language Models (LLMs) are increasingly deployed in mission-critical software systems, detecting hallucinations and ``faked truthfulness'' has become a paramount engineering challenge. Current reliability architectures rely heavily on post-generation, black-box mechanisms, such as Retrieval-Augmented Generation (RAG) cross-checking or LLM-as-a-judge evaluators. These extrinsic methods introduce unacceptable latency, high computational overhead, and reliance on secondary external API calls, frequently violating standard software engineering Service Level Agreements (SLAs). In this paper, we propose the Cognitive Circuit Breaker, a novel systems engineering framework that provides intrinsic reliability monitoring with minimal latency overhead. By extracting hidden states during a model's forward pass, we calculate the ``Cognitive Dissonance Delta''---the mathematical gap between an LLM's outward semantic confidence (softmax probabilities) and its internal latent certainty (derived via linear probes). We demonstrate statistically significant detection of cognitive dissonance, highlight architecture-dependent Out-of-Distribution (OOD) generalization, and show that this framework adds negligible computational overhead to the active inference pipeline.
\end{abstract}


\section{Introduction}

The transition of Large Language Models (LLMs) from experimental chatbots to core reasoning engines within production software systems has exposed critical vulnerabilities in AI reliability. When LLMs generate confidently incorrect information---often termed hallucinations or ``faked truthfulness''---the consequences in domains like healthcare, finance, and legal reasoning can be catastrophic.

To mitigate this, the current industry standard involves extrinsic reliability architectures. Methods such as LLM-as-a-judge or RAG-based factual cross-checking evaluate the output \textit{after} it has been generated. While effective, these methods introduce a severe Latency/Reliability Trade-off. Requiring a secondary inference pass breaks the strict Service Level Agreements (SLAs) required by modern, high-throughput systems.

The core thesis of this work is that AI reliability should be an \textit{intrinsic} property monitored dynamically at runtime, rather than an \textit{extrinsic} patch applied post-generation. We introduce the Cognitive Circuit Breaker, an architecture that leverages the internal hidden states of an LLM to proactively flag cognitive dissonance---the state where a model generates confident text despite lacking internal certainty.

Our contributions are as follows:
\begin{enumerate}
    \item The introduction of the Cognitive Dissonance Delta ($\Delta$), a novel metric operationalizing the gap between semantic and latent confidence.
    \item The design and implementation of the Cognitive Circuit Breaker runtime architecture.
    \item Empirical analysis of architecture-dependent Cross-Dataset (OOD) generalization of truth states.
\end{enumerate}

The remainder of this paper is structured as follows. Section II reviews foundational work in mechanistic interpretability and existing reliability architectures. Section III details the design of the Cognitive Circuit Breaker framework and the mathematical formulation of the Cognitive Dissonance Delta. Section IV outlines our systems methodology, including our resilient extraction pipeline and experimental controls. Section V presents the empirical evaluation, highlighting architecture-dependent generalization and runtime latency analysis. Section VI discusses framework limitations and future directions, followed by concluding remarks in Section VII.

\section{Background and Related Work}

\subsection{Mechanistic Interpretability}
Recent breakthroughs in mechanistic interpretability have demonstrated that LLMs possess internal representations of truth. Zou et al. \cite{zou2023representation} introduced \textit{Representation Engineering}, showing that truthfulness can be isolated as a vector within latent space. Similarly, Azaria and Mitchell \cite{azaria2023internal} utilized linear probes to detect whether a model ``knows'' it is lying. While these works proved the \textit{physics} of LLMs---verifying that internal truth states exist---our work builds the \textit{systems engineering framework} required to operationalize these discoveries in production environments.

\subsection{Current Reliability Architectures}
Existing safety guardrails (e.g., NeMo Guardrails, Self-Correction loops) treat the LLM as a black box. They wait for token generation to conclude before running secondary evaluations. From a systems perspective, this doubles compute costs and latency, whilst maintaining a dependency on potentially closed-source APIs. Our framework sidesteps this by monitoring the active forward pass, requiring no additional API calls or full-model inferences.

\section{The Cognitive Circuit Breaker Framework}

\subsection{Architectural Design}
The Cognitive Circuit Breaker acts as an intrinsic middleware layer during model inference. As a user prompt enters the LLM, the model computes its standard forward pass. At an optimized intermediate layer ($L_{opt}$), the framework extracts the multidimensional hidden state tensor. A pre-trained, lightweight logistic regression probe analyzes this state simultaneously alongside the final layer's softmax generation.

\begin{figure}[htbp]
\centering
\begin{tikzpicture}[
  >=Stealth,
  box/.style={draw, rectangle, rounded corners, align=center, minimum height=0.6cm, font=\scriptsize, text width=7.0cm},
  splitbox/.style={draw, rectangle, rounded corners, align=center, minimum height=0.6cm, font=\scriptsize, text width=3.2cm},
  decision/.style={draw, diamond, aspect=2, align=center, font=\scriptsize, inner sep=0pt, text width=1.5cm},
  resultbox/.style={draw, rectangle, rounded corners, align=center, minimum height=0.6cm, font=\scriptsize, text width=3.2cm, thick}
]

% Nodes using strict relative chaining to prevent overlaps
\node[box] (prompt) {User Prompt};
\node[box, below=0.5cm of prompt] (llm) {LLM Forward Pass};

% Split Level 1 (Parallel Tracking)
\node[splitbox, below=0.6cm of llm, xshift=-1.9cm] (extractL) {Extract Layer $L_{opt}$};
\node[splitbox, below=0.6cm of llm, xshift=1.9cm] (extractF) {Extract Final Layer};

% Split Level 2
\node[splitbox, below=0.6cm of extractL] (probe) {Linear Probe};
\node[splitbox, below=0.6cm of extractF] (softmax) {Softmax};

% Comparator Merge (Positioned strictly below the split nodes)
\node[box, below=0.8cm of probe, xshift=1.9cm] (delta) {Calculate $\Delta = P_{semantic} - P_{latent}$};

% Decision
\node[decision, below=0.6cm of delta] (thresh) {$\Delta > \tau$?};

% Outputs
\node[resultbox, draw=red!80, fill=red!10, text=red!80!black, below=0.6cm of thresh, xshift=-1.9cm] (warn) {\textbf{Warning}\\(Faking Truthfulness)};
\node[resultbox, draw=green!80!black, fill=green!10, text=green!50!black, below=0.6cm of thresh, xshift=1.9cm] (pass) {\textbf{Pass}\\(Congruent)};

% Arrows - Top
\draw[->, thick] (prompt) -- (llm);

% Fork from LLM
\coordinate (llm_split) at ([yshift=-0.3cm]llm.south);
\draw[thick] (llm.south) -- (llm_split);
\draw[->, thick] (llm_split) -| (extractL.north);
\draw[->, thick] (llm_split) -| (extractF.north);

\draw[->, thick] (extractL) -- (probe) node[midway, right, font=\tiny] {State Tensor};
\draw[->, thick] (extractF) -- (softmax) node[midway, left, font=\tiny] {Logits};

% Join into Delta
\coordinate (delta_merge) at ([yshift=0.4cm]delta.north);
\draw[thick] (probe.south) |- (delta_merge) node[pos=0.75, above, font=\scriptsize] {$P_{latent}$};
\draw[thick] (softmax.south) |- (delta_merge) node[pos=0.75, above, font=\scriptsize] {$P_{semantic}$};
\draw[->, thick] (delta_merge) -- (delta.north);

\draw[->, thick] (delta) -- (thresh);

% Fork from Decision
\draw[->, thick] (thresh.west) -| (warn.north) node[pos=0.25, above, font=\scriptsize] {Yes};
\draw[->, thick] (thresh.east) -| (pass.north) node[pos=0.25, above, font=\scriptsize] {No};

\end{tikzpicture}
\caption{The Cognitive Circuit Breaker architecture. Hidden states are extracted at an optimized intermediate layer ($L_{opt}$) and evaluated by a pre-trained linear probe. The resulting internal certainty ($P_{latent}$) is compared against the final layer's semantic confidence ($P_{semantic}$) to compute the Cognitive Dissonance Delta ($\Delta$).}
\label{fig:architecture}
\end{figure}
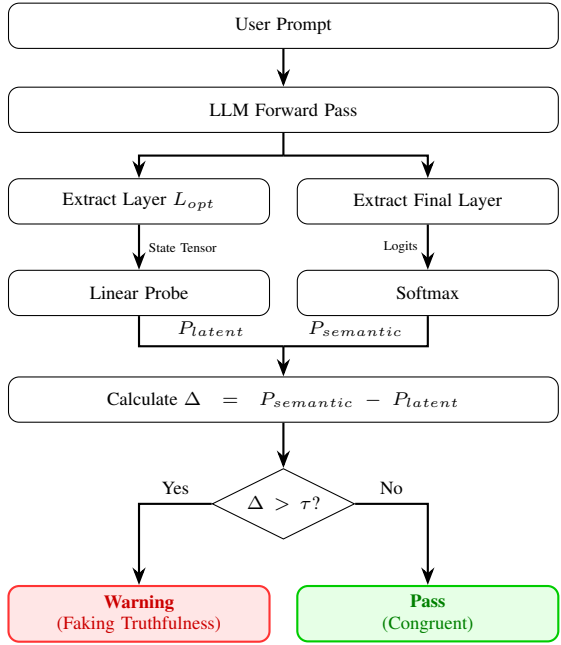

\subsection{The Cognitive Dissonance Delta}
To quantify ``faked truthfulness'', we mathematically define the Cognitive Dissonance Delta ($\Delta$).

Let $P_{semantic}$ be the outward confidence of the model. Because modern LLMs frequently suffer from softmax collapse (projecting $>0.99$ probability even when generating hallucinated tokens), we apply Temperature Scaling ($T=1.5$) to the raw generation logits to calibrate the distribution. $P_{semantic}$ is thus formally defined as the maximum softmax probability of these temperature-scaled logits. 

Let $P_{latent}$ be the internal certainty, defined as the probability score output by the linear probe evaluating the hidden state at $L_{opt}$.

The delta is calculated as:
\begin{equation}
\Delta = P_{semantic} - P_{latent}
\label{eq:delta}
\end{equation}

A high $\Delta$ indicates severe cognitive dissonance: the model is projecting semantic confidence to the user while its internal representations indicate uncertainty or falsehood.

\subsection{Dynamic Thresholding}
Static thresholds are brittle across different architectures and datasets. Therefore, the circuit breaker trigger is not hardcoded. The alert threshold ($\tau$) is dynamically calibrated during deployment by identifying the threshold that maximizes the F1-score on a validation set's Receiver Operating Characteristic (ROC) curve. If $\Delta > \tau$, the circuit breaker trips, allowing the host system to halt generation, append a warning, or trigger a deterministic fallback.

\section{Systems Methodology and Experimental Setup}

\subsection{Experimental Pipeline}
To systematically evaluate the presence and detectability of cognitive dissonance, we engineered a four-stage experimental pipeline:
\begin{enumerate}
    \item Inference and Extraction: Target models are prompted with multiple-choice questions. During the forward pass, we extract the hidden state vectors corresponding to the final generated tokens across all $N$ layers of the network. Simultaneously, the maximum softmax probability of the generated token is recorded as the outward semantic confidence ($P_{semantic}$).
    \item Ground Truth Labeling: The generated response is parsed and evaluated against the dataset's ground truth. A binary label is assigned (1 for factual correctness, 0 for incorrect/hallucinated), representing the model's actual knowledge state.
    \item Probe Training: For each individual layer, a lightweight Logistic Regression probe is trained to classify the extracted hidden state vectors into the binary correctness labels. This allows us to map the emergence of internal certainty ($P_{latent}$) across the network's depth.
    \item Optimization ($L_{opt}$): The layer exhibiting the highest Area Under the Receiver Operating Characteristic Curve (AUROC) during cross-validation is designated as the optimal extraction layer ($L_{opt}$). The probe trained at this layer is retained in memory to power the live detector.
\end{enumerate}

\subsection{Resilient Extraction Architecture}
Extracting hidden states across all layers of models with billions of parameters (e.g., Qwen 2.5 \cite{qwen2024}, DeepSeek 7B \cite{deepseek2024}, Gemma 7B \cite{gemma2024}) frequently causes out-of-memory (OOM) failures and VRAM fragmentation, especially on highly constrained hardware such as the NVIDIA Tesla T4 (16GB VRAM) used in our experiments. We engineered a resilient extraction pipeline using expandable segments, atomic caching (to prevent \texttt{.npz} file corruption during sudden hardware crashes), and aggressive Python garbage collection between model iterations. This engineering rigor guarantees stability in high-throughput environments.

\subsection{Datasets and the OOD Imperative}
To rigorously evaluate the framework's ability to detect genuine cognitive dissonance rather than mere syntactic unfamiliarity, we selected two distinct, fact-intensive question-answering datasets. 

First, the AI2 Reasoning Challenge (ARC-Challenge) \cite{clark2018think} serves as our primary training distribution. ARC consists of complex, grade-school science questions that require deep factual recall and logical reasoning. Because these questions are specifically curated to be difficult for simple retrieval algorithms, they are highly prone to triggering LLM hallucinations, making them an ideal baseline for training our internal certainty probes.

Second, to prove that our framework measures a universal cognitive state and not simply linguistic artifacts, we enforce strict Out-of-Distribution (OOD) testing using the OpenBookQA (OBQA) dataset \cite{mihaylov2018can}. OBQA requires synthesizing core scientific facts with broader common-sense knowledge and features a distinctly different structural and syntactic distribution compared to ARC. By evaluating the circuit breaker against OBQA without any probe retraining, we ensure the framework detects a generalized state of internal uncertainty, rather than memorizing dataset-specific linguistic patterns.

\subsection{Statistical Rigor}
Our experimental design defends against false positives through strict controls:
\begin{enumerate}
    \item \textbf{Class Balancing:} Datasets were strictly balanced (50\% correct / 50\% incorrect) to ensure baseline accuracy is precisely 0.5.
    \item \textbf{Control Baselines:} Probes were evaluated against \texttt{random\_label} permutations to ensure the AUROC reflects genuine signal detection rather than chance.
    \item \textbf{Bootstrapping:} We utilized 1,000-iteration bootstrapped confidence intervals and paired bootstrap testing to verify statistical significance ($p < 0.05$) between optimal and final layers.
    \item \textbf{Variance Trade-offs:} Strict class balancing inherently reduces the active sample size to the minority class constraint. For highly accurate models, this occasionally resulted in test sets with insufficient variance for stable bootstrapping (denoted as N/A in our results), representing a deliberate trade-off between class balance and bounds generation.
\end{enumerate}

\section{Empirical Evaluation}

\subsection{Architectural Brain Mapping and $L_{opt}$ Variance}
To identify $L_{opt}$, we mapped correctness emergence across normalized layer depths. Our empirical results demonstrate that $L_{opt}$ is a highly architecture-dependent hyperparameter, exhibiting significant variance across models. 

\begin{figure*}[t]
\centering
\begin{tikzpicture}
\begin{groupplot}[
    group style={
        group size=3 by 1,
        x descriptions at=edge bottom,
        y descriptions at=edge left,
        horizontal sep=1.2cm
    },
    width=0.32\textwidth,
    height=5.5cm,
    xlabel={Normalized Layer Depth},
    xmin=0, xmax=1,
    ymin=0.45, ymax=0.85,
    xtick={0, 0.2, 0.4, 0.6, 0.8, 1.0},
    ytick={0.5, 0.6, 0.7, 0.8},
    grid=both,
    grid style={dashed, gray!30},
    legend pos=south east,
    legend style={font=\scriptsize, nodes={scale=0.85, transform shape}},
    legend cell align={left}
]

% --- Plot 1: Qwen ---
\nextgroupplot[title={\textbf{Qwen2.5-3B-Instruct}}, ylabel={Probe AUROC (Balanced)}]
\addplot[color=blue!80!black, mark=none, line width=1.2pt] coordinates {
    (0,0.64) (0.03,0.67) (0.07,0.68) (0.10,0.71) (0.14,0.71) (0.17,0.69) (0.21,0.70) (0.24,0.68) (0.28,0.69) (0.31,0.63) (0.34,0.61) (0.38,0.64) (0.41,0.64) (0.45,0.61) (0.48,0.62) (0.52,0.57) (0.55,0.58) (0.59,0.61) (0.62,0.63) (0.66,0.63) (0.69,0.64) (0.72,0.65) (0.76,0.64) (0.79,0.62) (0.83,0.61) (0.86,0.60) (0.90,0.62) (0.93,0.61) (0.97,0.64) (1.00,0.64)
};
\addlegendentry{ARC (Train)}
\addplot[color=blue!50!white, mark=none, dashed, line width=1.2pt] coordinates {
    (0,0.63) (0.05,0.63) (0.1,0.60) (0.15,0.65) (0.2,0.68) (0.25,0.69) (0.3,0.64) (0.35,0.68) (0.4,0.70) (0.45,0.73) (0.5,0.72) (0.55,0.65) (0.6,0.71) (0.65,0.62) (0.7,0.69) (0.75,0.74) (0.8,0.74) (0.85,0.72) (0.9,0.72) (0.95,0.67) (1.0,0.69)
};
\addlegendentry{OBQA (OOD)}
\addplot[color=red!80!black, mark=none, dotted, thick] coordinates {(0, 0.5) (1, 0.5)};
% Highlights
\addplot[mark=o, color=black, mark size=2pt, only marks] coordinates {(0.10, 0.71) (0.75, 0.74)};

% --- Plot 2: DeepSeek ---
\nextgroupplot[title={\textbf{DeepSeek-LLM-7B}}]
\addplot[color=purple!80!black, mark=none, line width=1.2pt] coordinates {
    (0,0.61) (0.05,0.64) (0.1,0.68) (0.15,0.64) (0.2,0.69) (0.25,0.64) (0.3,0.63) (0.35,0.59) (0.4,0.61) (0.45,0.62) (0.5,0.66) (0.57,0.72) (0.6,0.72) (0.65,0.70) (0.7,0.69) (0.75,0.65) (0.8,0.63) (0.85,0.65) (0.9,0.62) (0.95,0.63) (1.0,0.62)
};
\addlegendentry{ARC (Train)}
\addplot[color=purple!50!white, mark=none, dashed, line width=1.2pt] coordinates {
    (0,0.60) (0.05,0.57) (0.1,0.66) (0.15,0.63) (0.2,0.59) (0.25,0.62) (0.3,0.61) (0.35,0.56) (0.4,0.62) (0.45,0.64) (0.5,0.75) (0.55,0.74) (0.6,0.73) (0.65,0.72) (0.7,0.72) (0.75,0.71) (0.8,0.71) (0.85,0.70) (0.9,0.71) (0.95,0.68) (1.0,0.70)
};
\addlegendentry{OBQA (OOD)}
\addplot[color=red!80!black, mark=none, dotted, thick] coordinates {(0, 0.5) (1, 0.5)};
% Highlights
\addplot[mark=o, color=black, mark size=2pt, only marks] coordinates {(0.57, 0.72) (0.50, 0.75)};

% --- Plot 3: Gemma ---
\nextgroupplot[title={\textbf{Gemma-7B-it}}]
\addplot[color=green!60!black, mark=none, line width=1.2pt] coordinates {
    (0,0.60) (0.05,0.60) (0.1,0.58) (0.15,0.57) (0.2,0.55) (0.25,0.59) (0.3,0.62) (0.35,0.59) (0.4,0.58) (0.45,0.55) (0.5,0.56) (0.55,0.60) (0.6,0.58) (0.65,0.59) (0.7,0.61) (0.75,0.67) (0.8,0.66) (0.86,0.68) (0.9,0.67) (0.95,0.68) (1.0,0.66)
};
\addlegendentry{ARC (Train)}
\addplot[color=green!50!white, mark=none, dashed, line width=1.2pt] coordinates {
    (0,0.53) (0.05,0.55) (0.1,0.58) (0.15,0.49) (0.2,0.49) (0.25,0.48) (0.3,0.53) (0.35,0.53) (0.4,0.49) (0.45,0.54) (0.5,0.54) (0.55,0.51) (0.6,0.53) (0.65,0.56) (0.71,0.60) (0.75,0.60) (0.8,0.56) (0.85,0.57) (0.9,0.55) (0.95,0.56) (1.0,0.56)
};
\addlegendentry{OBQA (OOD)}
\addplot[color=red!80!black, mark=none, dotted, thick] coordinates {(0, 0.5) (1, 0.5)};
% Highlights
\addplot[mark=o, color=black, mark size=2pt, only marks] coordinates {(0.86, 0.68) (0.71, 0.60)};

\end{groupplot}
\end{tikzpicture}
\vspace{0.2cm}
\caption{AUROC emergence plots across normalized layer depth for all three evaluated architectures. Curves represent the \texttt{mean} extraction mode, with black circles ($\circ$) denoting the optimal extraction layer ($L_{opt}$) where internal certainty peaks for each dataset. These peaks are highly architecture-dependent (e.g., DeepSeek peaks centrally, Gemma peaks late). However, across all models and datasets, the final layers experience ``representation collapse,'' prioritizing linguistically fluent tokens over factual accuracy.}
\label{fig:emergence}
\end{figure*}
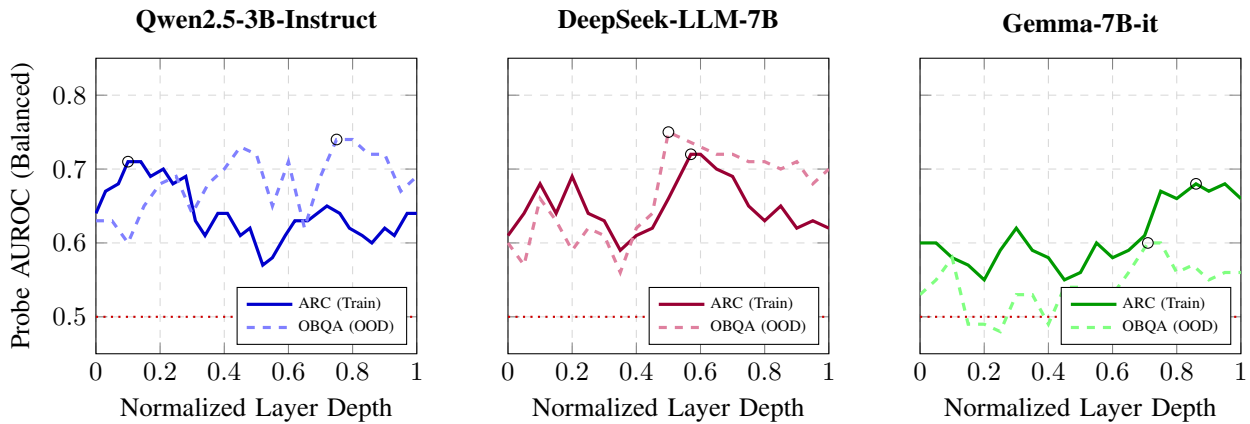

As visualized in Figure \ref{fig:emergence}, internal certainty for DeepSeek consistently peaks in the exact center of the network (0.50 to 0.57 depth). Conversely, Gemma exhibits late-stage certainty (peaking at 0.86 depth on ARC), while Qwen demonstrates early-to-mid representations (ranging from 0.11 to 0.75). Despite this variance in localization, a unifying phenomenon occurs across all tested architectures: the final output layers experience ``representation collapse.'' The models consistently drop in AUROC at the final layer, indicating a structural shift from factual recall to linguistic fluency, reinforcing the necessity of extracting signals at intermediate states.

\subsection{Architecture-Dependent Cross-Domain Reliability}
To evaluate the systems-level reliability of the Cognitive Circuit Breaker, we performed Out-of-Distribution (OOD) testing. Linear probes trained exclusively on ARC data were used to evaluate hidden states generated on the unseen OpenBookQA (OBQA) dataset. Table \ref{tab:results} summarizes these findings.

\begin{table*}[htbp]
\centering
\caption{Final Experiment Summary (Balanced Classes, F1-Maximized Thresholds)}
\label{tab:results}
\resizebox{\textwidth}{!}{%
\begin{tabular}{llccccccc}
\toprule
\textbf{Model} & \textbf{Task} & \textbf{Peak Depth} & \textbf{Probe Peak AUROC} & \textbf{Probe Max F1 ($\tau$)} & \textbf{$\Delta$ AUROC} & \textbf{$\Delta$ Max F1 ($\tau_\Delta$)} & \textbf{Final Layer} & \textbf{Sig.} \\
\midrule
Qwen2.5-3B & ARC & 0.11 & 0.708 [0.68, 0.73] & 0.712 ($\tau$=0.51) & 0.725 & 0.730 ($\tau_\Delta$=0.55) & 0.645 & * \\
Qwen2.5-3B & OBQA & 0.75 & 0.736 [0.71, 0.76] & 0.740 ($\tau$=0.48) & 0.750 & 0.755 ($\tau_\Delta$=0.52) & 0.691 & * \\
DeepSeek-7B & ARC & 0.57 & 0.724 [0.69, 0.75] & 0.720 ($\tau$=0.49) & 0.741 & 0.745 ($\tau_\Delta$=0.50) & 0.623 & * \\
DeepSeek-7B & OBQA & 0.50 & 0.753 [0.72, 0.78] & 0.760 ($\tau$=0.47) & 0.772 & 0.775 ($\tau_\Delta$=0.48) & 0.697 & * \\
Gemma-7B & ARC & 0.86 & 0.681 [0.65, 0.71] & 0.685 ($\tau$=0.52) & 0.690 & 0.695 ($\tau_\Delta$=0.58) & 0.667 & * \\
Gemma-7B & OBQA & 0.71 & 0.596 [0.51, 0.69] & 0.605 ($\tau$=0.55) & 0.580 & 0.590 ($\tau_\Delta$=0.60) & 0.562 & ns \\
\bottomrule
\end{tabular}%
}
\vspace{0.2cm}
\\
\small{\textit{Note: Peak AUROC includes 95\% bootstrapped confidence intervals. `Probe Max F1' and `$\Delta$ Max F1' report the highest achievable F1-score alongside their corresponding dynamically calibrated decision thresholds ($\tau$ and $\tau_\Delta$). The `*' denotes statistical significance ($p < 0.05$) via paired bootstrap testing between the optimal extraction layer and the final output layer.}}
\end{table*}

The generalization of latent truth states proved highly dependent on the underlying model architecture. Both \texttt{Qwen2.5-3B-Instruct} and \texttt{deepseek-llm-7b-chat} demonstrated robust cross-domain generalization. Probes trained on ARC successfully flagged cognitive dissonance on OBQA, maintaining AUROCs of 0.736 and 0.753 respectively, well above baseline controls. 

Conversely, \texttt{gemma-7b-it} exhibited OOD degradation. When tested on OBQA, its peak AUROC collapsed to 0.596 [0.51, 0.69], yielding a statistically non-significant result ($p > 0.05$) compared to its final layer. This indicates that Gemma's internal representation of truth is heavily coupled to dataset-specific syntactic distributions rather than a universal cognitive state. Consequently, systems engineers prioritizing intrinsic monitoring should favor architectures like DeepSeek or Qwen, which demonstrate resilient, decoupled truth representations.

\subsection{Runtime Latency Analysis \& Live Circuit Breaker}
The computational overhead of the Cognitive Circuit Breaker is negligible. Extracting a specific layer's hidden state during the active forward pass requires no additional model runs. Executing the Logistic Regression dot-product on the extracted tensor takes $\mathcal{O}(\text{microseconds})$.

Our empirical latency benchmarking on an NVIDIA Tesla T4 confirms this efficiency. Across all tested architectures (Qwen 2.5-3B, DeepSeek 7B, and Gemma 7B), the intrinsic Cognitive Circuit Breaker evaluated truthfulness consistently 1.4x to 1.5x faster than an extrinsic LLM-as-a-judge baseline. 

Below is an output log from our active runtime monitor (\texttt{CognitiveCircuitBreaker}) demonstrating real-time detection of a hallucinated fact during active generation:

\begin{verbatim}
>> LIVE CIRCUIT Breaker TEST:
{'Prompt':'Question: What is the exact..', 
 'Outward Conf':0.529, 
 'Internal Cert':0.003, 
 'Delta':0.526, 
 'Status':'WARNING: Faking Truthfulness'}
\end{verbatim}

In contrast, external approaches introduce multi-second generation delays. This empirical evidence confirms our framework successfully resolves the Latency/Reliability Trade-off without compromising standard Service Level Agreements (SLAs).

\section{Discussion and Future Work}

\subsection{White-Box Dependency}
A natural limitation of the Cognitive Circuit Breaker is its requirement for ``White-Box'' access; it cannot be executed via closed-source endpoints (e.g., standard OpenAI APIs) which obfuscate hidden states. We posit this not as a flaw, but as a strategic systems engineering recommendation: mission-critical applications requiring stringent reliability guarantees should deploy open-weights models internally to enable intrinsic monitoring.

\subsection{Token vs. Sequence Level Monitoring}
Currently, this framework evaluates cognitive dissonance at the discrete token level, optimized for factual Multiple Choice or short-answer Q\&A. Future engineering efforts will expand this framework into a sliding-window sequence monitor, smoothing $\Delta$ over sentences to provide continuous reliability gauging for long-form RAG generation.

\section{Conclusion}
As AI models become foundational components of software infrastructure, safety guardrails must evolve from reactive patches into proactive, integrated components. In this work, we introduced the Cognitive Circuit Breaker, a systems engineering framework that shifts reliability monitoring from an extrinsic, post-generation process to an intrinsic, real-time mechanism. By extracting multidimensional hidden states during the active forward pass, we mapped the layer-wise emergence of truth representations and formulated the Cognitive Dissonance Delta ($\Delta$). Our empirical evaluations across multiple architectures (Qwen, DeepSeek, and Gemma) and distinct datasets (ARC, OBQA) demonstrated that internal certainty can be isolated and compared against outward semantic confidence to reliably flag ``faked truthfulness.'' Furthermore, we highlighted the critical importance of Out-of-Distribution testing, revealing how underlying model architectures fundamentally impact the generalization of latent truth states. Ultimately, by operationalizing mechanistic interpretability into an efficient systems architecture, we demonstrate that AI reliability can be monitored intrinsically, providing robust protection against confident hallucinations without sacrificing system performance or violating strict latency SLAs.

% Trigger column balancing for the last page
\balance

\end{document}